\documentclass[12pt]{scrartcl}
\usepackage{authblk}

\usepackage[applemac]{inputenc}
\usepackage{latexsym}
\usepackage{amssymb,amsmath,amsthm}

\usepackage{graphicx}
\usepackage{color,colortbl}
\usepackage{tikz}
\usetikzlibrary{shapes.geometric}
\usetikzlibrary{arrows,shapes,backgrounds}
\usepackage{enumerate}
\usepackage{wrapfig}
\usepackage{algorithm}
\usepackage{algorithmic}
\usepackage[toc,page]{appendix}

\usepackage[numbers,square,comma]{natbib} 

\def\citeyear{\cite}

\definecolor{shade}{RGB}{241, 240, 247}

\begin{document}

\def\Nset{\mathbb{N}}
\def\Ascr{\mathcal{A}}
\def\Bscr{\mathcal{B}}
\def\Cscr{\mathcal{C}}
\def\Dscr{\mathcal{D}}
\def\Escr{\mathcal{E}}
\def\Fscr{\mathcal{F}}
\def\Hscr{\mathcal{H}}
\def\Iscr{\mathcal{I}}
\def\Mscr{\mathcal{M}}
\def\Nscr{\mathcal{N}}
\def\Pscr{\mathcal{P}}
\def\Qscr{\mathcal{Q}}
\def\Rscr{\mathcal{R}}
\def\Sscr{\mathcal{S}}
\def\Vscr{\mathcal{V}}
\def\Wscr{\mathcal{W}}
\def\Xscr{\mathcal{X}}
\def\cupp{\stackrel{.}{\cup}}
\def\bold{\bf\boldmath}
\def\A{\ensuremath{\mathcal{A}}}

\newcommand{\rouge}[1]{\textcolor{red}{\tt \footnotesize #1}}
\newcommand{\boldheader}[1]{\smallskip\noindent{\bold #1:}\quad}
\newcommand{\caseclaim}[1]{\smallskip\noindent{\textbf{#1:}}}
\newcommand{\PP}{\mbox{\slshape P}}
\newcommand{\NP}{\mbox{\slshape NP}}
\newcommand{\MAXSNP}{\mbox{\slshape MAXSNP}}
\newcommand{\opt}{\mbox{$\small\mathrm{OPT}$}}
\newcommand{\mst}{\mbox{$\small\mathrm{MST}$}}
\newcommand{\delay}{{\mathrm{delay}}}
\newcommand{\level}{{$\mathrm{level}$}}
\newcommand{\lp}{\mbox{\scriptsize\mathrm{$LP$}}}
\newtheorem{theorem}{Theorem}
\newtheorem{lemma}[theorem]{Lemma}
\newtheorem{corollary}[theorem]{Corollary}
\newtheorem{proposition}[theorem]{Proposition}
\newtheorem{definition}[theorem]{Definition}
\newtheorem{remark}[theorem]{Remark}
\def\prove{\par \noindent \hbox{\textbf{Proof:}}\quad}
\def\endproof{\eol \rightline{$\Box$} \par}
\renewcommand{\endproof}{\hspace*{\fill} {\boldmath $\Box$} \par \vskip0.5em}
\newcommand{\mathendproof}{\vskip-1.8em\hspace*{\fill} {\boldmath $\Box$} \par \vskip1.8em}
\newcommand{\sfrac}[2]{{\textstyle{\frac{#1}{#2}}}}

\newcommand{\odd}{\mbox{\rm odd}}
\newcommand{\revalg}{Algorithm \ref{alg:bicrit-neu}b}

\definecolor{orange}{rgb}{1,0.9,0}
\definecolor{violet}{rgb}{0.8,0,1}
\definecolor{darkgreen}{rgb}{0,0.5,0}
\definecolor{grey}{rgb}{0.75,0.75,0.75}

\newcommand{\jnote}[1]{[{JK: \textcolor{red}{#1}}]\marginpar{\includegraphics[width=2ex]{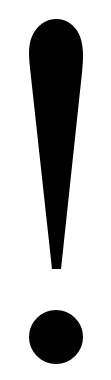}}}
\newcommand{\snote}[1]{[{SH: \textcolor{red}{#1}}]\marginpar{\includegraphics[width=2ex]{exclamation_mark.png}}}
\newcommand{\jvnote}[1]{[{JV: \textcolor{red}{#1}}]\marginpar{\includegraphics[width=2ex]{exclamation_mark.png}}}

\title{Vehicle Routing with Subtours} 

\author[1]{Stephan Held}
\author[2]{Jochen Könemann}
\author[1]{Jens Vygen}
\affil[1]{\normalsize Research Institute for Discrete Mathematics, University of Bonn}
\affil[2]{Department of Combinatorics \& Optimization, University of Waterloo}
\affil[ ]{held@dm.uni-bonn.de, jochen@uwaterloo.ca, vygen@dm.uni-bonn.de}
\renewcommand\Authands{ and }
   


\date{\today}



\maketitle

\begin{abstract}
When delivering items to a set of destinations, one can save time and cost
by passing a subset to a sub-contractor at any point en route. 
We consider a model where a set of items
are initially loaded in one vehicle and should be distributed before a given deadline $\Delta$.
In addition to travel time and time for deliveries, we assume that there is a fixed delay for handing
over an item from one vehicle to another.

We will show that it is easy to decide whether an instance is
feasible, i.e., whether it is possible to deliver all items before
the deadline $\Delta$. We then consider computing a feasible tour of
minimum cost, where we incur a cost per unit distance traveled by the
vehicles, and a setup cost for every used vehicle. 
Our problem arises in practical applications and generalizes classical problems such as
shallow-light trees and the bounded-latency problem.

Our main result is a polynomial-time algorithm that, for
any given $\epsilon>0$ and any feasible instance, computes a solution
that delivers all items before time $(1+\epsilon)\Delta$ and has
cost  $\mathcal{O}(1+\frac{1}{\epsilon})\opt$, where $\opt$ is the minimum
cost of any feasible solution.

Known algorithms for special cases begin with a cheap solution and 
decompose it where the deadline is violated.
This alone is insufficient for our problem.
Instead, we also need a fast solution to start with, and 
a key feature of our algorithm is a careful combination of cheap and fast solutions.
We show that our result is best possible in the sense that any improvement
would lead to progress on 25-year-old questions on shallow-light trees. 

\end{abstract}

\section{Introduction}

Logistics companies are exploring innovative ways of delivering items
to customers.  In particular with an increasing number of crowdsourced
delivery startups \cite{RM14} comes new flexibility in designing
delivery routes; e.g., \cite{KZB17} studies delivery applications
where crowdsourcing is used to facilitate last-leg delivery of items.
In the classical setting, a set of vehicles are initially located at a
central depot that in turn serves as the starting point for all tours
serving customers. Where deadlines are tight, this model naturally
leads to a large number of tours, and implementing these solutions
necessitates the maintenance of large vehicle fleets. 

Our work is motivated by instances where the depot is relatively far
from many customers. In designing delivery schedules, one would
therefore want to use bigger vehicles to transport a large set of
items closer to clusters of customers at which point one would then
utilize smaller vehicles for the final leg of the delivery process.  To
model this situation, we introduce a new kind of vehicle routing
problem and study its approximability.

Let us assume that a set of items is initially loaded on one vehicle
and that they need to be delivered to their destinations by a given
deadline $\Delta$.  The vehicle can deliver items itself, but it can
also -- at any place en route -- hand over a subset of its items to
another vehicle. This vehicle can then deliver these items or can again
hand over subsets to new vehicles en route.

Besides the normal travel cost per unit distance, we assume a fixed
setup cost for every vehicle that we use.  This assumption is of
course a simplification, but some companies have vehicles at many
different locations (almost everywhere) or hire sub-contractors that
are not paid for their way to the meeting point where items are handed
over to him.  For the same reason, the way back home is not paid;
therefore our tours are paths and not circuits.  We remark that
essentially the same problem arises when collecting items (or
students) and transporting them to a central location (or school).
However, to avoid confusion, we will speak of the delivery problem
only.


We do not consider vehicle capacities here, although implicitly the number of items that a vehicle can handle is bounded by the deadline.
We not only account for the travel time and the time to deliver items, but also consider a hand-over time proportional to the number of
items exchanged. 

Assuming that items can only be handed over from a vehicle to one
other vehicle simultaneously, then the resulting route structure can
best be described as an arborescence with maximum out-degree two,
where the root corresponds to the starting point of the initial vehicle, vertices with
out-degree two correspond to a hand-over, and all other vertices
correspond to a delivery.  We note that the imposed bound on the
out-degree of our vertices is not restrictive, as we allow placing
multiple vertices of the arborescence at the same geographical
location. 

\subsection{Problem definition}

Let us now describe the problem formally and introduce our notation.
An \textbf{instance} consists of a finite set $P$ of item destinations, a
root $r\notin P$, and a metric space $(M,c)$. We are also given a map
$\mu:\{r\}\cup P\to M$ that assigns root and items in $P$ to
locations in the metric space. Finally, we are given non-negative
parameters $\delta$, $\sigma$ and $\Delta$, capturing delivery time,
setup cost, and deadline, respectively.

We represent a \textbf{schedule} by an arborescence $(W,A)$ rooted at $r$ with $P\subset W$ and $\mu:W\setminus(P\cup\{r\})\to M$.
An arc $(x,y)\in A$ means that a vehicle travels from $x$ to $y$, causing delay $c(\mu(x),\mu(y))$.
At a vertex $p\in P$ we deliver the item $p$ and incur delay $\delta$.
At a vertex $w\in W\setminus P$ with out-degree 2 we split off a subtour and incur a hand-over delay
equal to the number of items handed over.
Note that it is always better to hand over at most half of the items that are currently in the vehicle.
We disallow out-degree greater than 2 because we need to specify the order of hand-overs.
Assuming that a vehicle cannot  simultaneously be involved in a  delivery and a hand-over,
we also forbid out-degree 2 for vertices in $P$.
However, we allow  that multiple vertices are mapped to the same point in the metric space, so multiple subtours can start at
the same point in the metric space.
A schedule allows that  items are handed over multiple times on their path from the root to their destination.

The \textbf{delay} of a schedule is the maximum delay to an item.
The \textbf{cost} of a schedule is $c(A,\mu)+s(W,A)$, where 
$c(A,\mu):=\sum_{(x,y)\in A} c(\mu(x),\mu(y))$ is the travel cost and
$s(W,A):=\sigma \cdot  |W_0|$ 
is the setup cost for  vehicles; here $W_0$ denotes the set of leaves in $(W,A)$. 
The goal is to find a minimum-cost schedule with delay at most $\Delta$. 



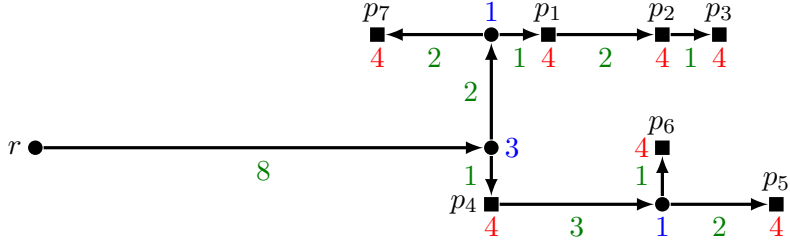
\begin{figure}
\begin{center}
\tikzset{>=latex}
\begin{tikzpicture}[scale=1.5, very thick, minimum size = 5, inner sep = 3,outer sep=2,square/.style={regular polygon,regular polygon sides=4}]
\begin{scope}[shift={(0,0)}]
	\node[draw,circle,fill, scale=0.5] (r) at (-4,0.5) {};       \node[left]  at (r)  {\small $r$};
	\node[draw,square,fill, scale=0.5] (p1) at (0.5,1.5) {};      \node[above]  at (p1)  {\small $p_1$}; \node[below] at (p1) {\textcolor{red}{\small $4$}};
	\node[draw,square,fill, scale=0.5] (p2) at (1.5,1.5) {};      \node[above]  at (p2)  {\small $p_2$}; \node[below] at (p2) {\textcolor{red}{\small $4$}};
	\node[draw,square,fill, scale=0.5] (p3) at (2,1.5) {};      \node[above]  at (p3)  {\small $p_3$}; \node[below] at (p3) {\textcolor{red}{\small $4$}};
	\node[draw,square,fill, scale=0.5] (p7) at (-1,1.5) {};    \node[above]  at (p7)  {\small $p_7$}; \node[below] at (p7) {\textcolor{red}{\small $4$}};

	\node[draw,square,fill, scale=0.5] (p4) at (0,0) {};   \node[left]  at (p4)  {\small $p_4$}; \node[below]  at (p4)  {\small \textcolor{red}{\small $4$}}; 
	\node[draw,square,fill, scale=0.5] (p5) at (2.5,0) {};   \node[above]   at (p5)  {\small $p_5$};\node[below]  at (p5)  {\small \textcolor{red}{\small $4$}}; 
	\node[draw,square,fill, scale=0.5] (p6) at (1.5,0.5) {};   \node[above]   at (p6)  {\small $p_6$};\node[left]  at (p6)  {\small \textcolor{red}{\small $4$}}; 
	\node[draw,circle,fill, scale=0.5] (s1) at (0,0.5) {}; \node[right] at (s1) {\textcolor{blue}{\small $3$}};

        \draw[->] (r)->(s1) node [below,midway] {\small\textcolor{darkgreen}{8}};
        \node[draw,circle,fill, scale=0.5] (s2) at (0,1.5) {};\node[above] at (s2) {\textcolor{blue}{\small $1$}};
        \draw[->] (s1)->(s2) node [left,midway] {\small\textcolor{darkgreen}{$2$}};
        \draw[->] (s2)->(p1) node [below,midway] {\small\textcolor{darkgreen}{$1$}};
        \draw[->] (p1)->(p2) node [below,midway] {\small\textcolor{darkgreen}{$2$}};
        \draw[->] (p2)->(p3) node [below,midway] {\small\textcolor{darkgreen}{$1$}};
        \draw[->] (s2)->(p7) node [below,midway] {\small\textcolor{darkgreen}{$2$}};
	\node[draw,circle,fill, scale=0.5] (s3) at (1.5,0) {};\node[below] at (s3) {\textcolor{blue}{\small $1$}};        
        \draw[->] (s1)->(p4) node [left,midway] {\small\textcolor{darkgreen}{$1$}};
        \draw[->] (p4)->(s3) node [below,midway] {\small\textcolor{darkgreen}{$3$}};
        \draw[->] (s3)->(p5) node [below,midway] {\small\textcolor{darkgreen}{$2$}};
        \draw[->] (s3)->(p6) node [left,midway] {\small\textcolor{darkgreen}{$1$}};

\end{scope}
\end{tikzpicture}
\end{center}
\caption{Example of a schedule. Edges are labeled by their distance (green). 
Vertices are labeled by hand-over delay (blue) or delivery delay (red, $\delta=4$).
The initial vehicle delivers $p_1$, $p_2$ and $p_3$. It hands over parcels $p_4$, $p_5$ and $p_6$ to a second vehicle
and later $p_7$ to a third vehicle. The second vehicle hands over $p_6$ to a fourth vehicle en route.
The delay of this schedule is $30$ (attained at $p_3$). Its cost is $23+4\sigma$.
\label{fig:example}}
\end{figure}

Figure \ref{fig:example} shows an example with seven items and a schedule with four vehicles.

\subsection{Overview of results and techniques}

As we will see in Section \ref{sec:relatedwork}, our problem is a common generalization of 
several problems studied in the literature, including shallow-light trees and the bounded-latency problem.
However, the possibility of starting subtours far away from the root and the handover delays make the problem more difficult.

Many previous approaches for special cases and similar problems, 
including the best-known algorithms for shallow-light trees and the bounded-latency problem (cf.\ Section \ref{sec:relatedwork}),
begin with a \emph{cheap} solution (minimum spanning tree or short TSP tour), traverse it, 
and split the tree or tour whenever necessary to make the solution fast enough, connecting the next vertex directly to the root.
However, this approach fails for our problem because connecting to the root can be too expensive and the 
order in which we split off subtours matters due to the handover delays.

We still use this {\em tour-splitting technique} to group the items, but in addition we need to
begin with a \emph{fast} solution.
Therefore we first show that there always exists a fastest schedule with 
\emph{caterpillar} structure, i.e., where all deliveries occur at leaves and the subgraph induced by the
vertices with out-degree 2 is a path. 
As a corollary, one can easily compute such a solution and
check whether a given instance is feasible (cf.\ Section \ref{sec:feasibility}). 

In Section \ref{sec:algorithm} we describe our algorithm.
It first uses the tour-splitting technique to partition the items into groups each of which is spanned by a short path and does not contain too many items. 
Next, our algorithm constructs a \emph{two-level caterpillar} so
that items in the same group are consecutive, without violating the
deadline much.  Although this helps saving length, this schedule still
uses a separate vehicle for every item and thus is usually very
expensive.  In a final step, we cluster subsets of groups to reduce the number of vehicles.

We compare our solution to a lower bound, composed of
the length of a minimum cost tree spanning $\{r\}\cup P$ (in the following denoted by $\mst$)
and a lower bound on the number of required vehicles.
This will yield our main result:

\begin{theorem}
\label{main}
Given a feasible instance and $\epsilon>0$, we can compute a
solution with delay at most $(1+\epsilon)\Delta$ and cost
$\mathcal{O}(1+\frac{1}{\epsilon})\opt$ in polynomial time, where $\opt$ is the minimum
 cost of a feasible schedule.
\end{theorem}

In Section \ref{sec:example} we will give an example that the tradeoff in Theorem 1 is unavoidable 
unless we use a significantly stronger lower bound.
In fact, the example also shows that the 25-year old result of Cong et al.~\citeyear{CK+92} on shallow-light trees
is best possible. Since this is a special case of our problem, 
improving the dependency on $\epsilon$  by more than a constant factor would require 
new lower bounds for shallow-light trees and immediately lead to progress 
on this very well-studied problem.

\subsection{Comments on our model}

Before we move on, let us comment on two subtle assumptions that we
made in our model, and argue why they are reasonable and necessary to
obtain such a result.

First, we assume $\delta\ge 1$, that is, delivering an item to its
final destination takes at least as long as handing it over to another
vehicle (we assumed the hand-over time to be 1 per item, which 
is of course no loss of generality by scaling).
This is certainly realistic in practice.  Some assumption on
$\delta$ is also necessary for our main result: with $\delta =0$,
there is no polynomial-time algorithm delivering all items of a
feasible instance before time $(1+\epsilon)\Delta$ for arbitrary small
$\epsilon>0$ (unless $\PP=\NP$).  To see this, scale down the
distances so that the shortest path beginning in $\mu(r)$ and
containing $\mu(p)$ for all $p\in P$ has length less than 1.  Then the
earliest deadline that we can meet is the length of a shortest such
path.  However, determining this length is APX-hard (by
straightforward reduction from the classical TSP \cite{PY93}; cf.\
Appendix \ref{sec:tsppathapxhard}).

Another subtle point in our model is that we pay $\sigma$ for every
vehicle, {\em including the initial one}.  Again, this looks
reasonable from the practical point of view, although one might also
think of the setup cost of the initial vehicle as already paid.  But
this assumption too is necessary for our main result.  If we had
assumed the initial vehicle to be free, then a very large value of
$\sigma$ would force any reasonably cheap solution to be a path, and
finding a path almost meeting a deadline is then equivalent to finding
an almost shortest tour starting at $r$ and visiting all item
destinations.  As above, this problem is APX-hard.

Finally, our tours and subtours are paths rather than circuits. 
This is not very important though, because if every vehicle had to return to its starting point,
the cost can at most double.


\subsection{Related work}
\label{sec:relatedwork}


The problem discussed in this paper generalizes
several well-studied, classical network design problems.  
For example, our problem contains the Steiner tree problem
(set $\sigma=0$ and $\Delta=\infty$) and is thus APX-hard \cite{CC02}.

An interesting special case of our problem arises when
$\sigma=0$ and $c$ is large compared to
$\delta$. In other words, the traveled distance dominates the delay
and the cost of any schedule. In this case, the goal would be to
compute a Steiner tree (or if $M=\mu(\{r\}\cup P)$ a spanning tree) 
that balances cost and the maximum distance from the root. 

Awerbuch et al.~\citeyear{ABP90} first showed that every finite metric space contains
a spanning tree whose diameter is at most a constant
times that of the underlying metric space, and whose weight is at most
a constant times that of a minimum-cost spanning tree.  
Such trees are called \emph{shallow-light trees}.
Cong et al.~\citeyear{CK+92} improved these results; they showed how to find, for any $\epsilon>0$, a 
spanning tree of length at most $(1+\frac{2}{\epsilon})\mst$ in which the 
path from $r$ to any other vertex is no longer than $1+\epsilon$ times
the maximum distance from $r$. 
Khuller, Raghavachari and Young~\citeyear{KRY95} generalized this
work to obtain, for any $\epsilon>0$, a tree $T$ with total cost at most $(1+\frac{2}{\epsilon})\mst$
such that, for every $v \in P$, the distance in $T$ from $r$ to $v$ is
at most $1+\epsilon$ times the distance between $r$ and $v$ in the metric space.

Khuller et al.~\citeyear{KRY95} also gave 
an example showing that the obtained tradeoff is best possible.
In Section \ref{sec:example} we generalize their example to prove that this is true even for instances
where all vertices have the same distance from $r$; implying that the result of \cite{CK+92} on shallow-light trees is also best possible.
This will also show that Theorem \ref{main} is
best possible unless we use a stronger lower bound than $\mst$.

A further generalization was given by Held and Rotter \citeyear{HR13},
considering Steiner trees and having 
an additional distance penalty per bifurcation.
However, in all of the  above algorithmic variants the result
may have many leaves and the delay models differ substantially from 
hand-over delays in our model.


There has been a tremendous amount of work on solving optimization
problems arising in the general context of vehicle routing, and we
cannot provide an adequate survey here.  We focus on the most
closely related work that we are aware of, and refer the reader to
Toth and Vigo's book~\citeyear{TV14} for a more
comprehensive introduction. 

Another interesting special case of our problem arises when delays are dominated by
travel times and cost is dominated by the setup cost.  Then it does not harm to start all subtours at
the position of the root, and the problem reduces to covering the items by as few paths as possible,
each starting at the root and having length at most $\Delta$.
Jothi and Raghavachari \citeyear{JR07} called this problem the \emph{bounded-latency problem}.
They observed that the tour-splitting technique
yields a solution violating the deadline by at most a factor of $(1+\epsilon)$ and using
at most $\frac{2}{\epsilon}$ times the optimum number of paths.

A similar problem is  the {\em distance-constrained vehicle routing problem} (DVRP), where
the goal is to cover the items by a minimum number of closed tours (returning to the root), each having  length at most $\Delta$.
Khuller, Malekian, and Mestre \citeyear{KMM11} and independently
Nagarajan and Ravi \citeyear{NR12} gave an  $(1+\epsilon, \mathcal{O}(\log\frac{1}{\epsilon}))$-bicriteria approximation algorithm.
This also works for the bounded-latency problem: 
partition the items according to their distance from $r$: 
items at distance more than $(1-2\epsilon)\Delta$ are in group 1, and
items at distance between $(1-2^j\epsilon)\Delta$ and $(1-2^{j-1}\epsilon)\Delta$ are in group $j$ 
($j=2,\ldots,\lceil\log\frac{1}{\epsilon}\rceil$).
Then each group $j$ is covered by (unrooted) paths, each of which has length at most $2^{j-1}\epsilon \Delta$
and can be completed by an edge to $r$, exceeding length $\Delta$ by at most $\epsilon \Delta$ (only in group 1).
The number of these paths can be minimized up to a factor 3 using an algorithm of \cite{AHL06}.
If $\opt$ is the number of tours in an optimal solution, we can cover each group by $2\opt$ paths
(shortcutting those that contain at least one item this group and splitting it into two if necessary).
Thus we end up with at most $6\opt$ paths in each group, and $6\lceil\log\frac{1}{\epsilon}\rceil\opt$ paths in total.

Connecting all paths to the root can, however, be much more expensive than splitting off subtours elsewhere:
for instance, if all items are to be delivered at the same position far away from the root, and the (relaxed) deadline prevents any tour from 
delivering more than one item, then the total length increases by a factor $|P|$ if we insist that
all paths start at the root. See also Appendix \ref{steiner-appendix} for a similar example.
 

Friggstad and Swamy \citeyear{FS14} studied regret-bounded variants
of vehicle routing problems and provided an
$\mathcal{O}(\log \Delta/ \log\log \Delta)$ approximation for DVRP
under the assumption  that the minimum distance in the underlying metric is at least one.
G{\o}rtz et al.~\citeyear{GM+16} considered various vehicle routing
problems in the setting where vehicles have non-uniform speeds and
capacities. Among other things, the authors study the variant of
DVRP where the vehicles 
have finite capacity and non-uniform speeds,  and
where the goal is to minimize the deadline.
G{\o}rtz et al.\ provide a
constant-factor approximation algorithm for this problem. 

Closely related to DVRP are vehicle routing problems with {\em
  min-max} objective. A typical such problem is the {\em min-max
  $X$-cover} problem, where
$X \in \{\mbox{path, tree, tour, }\ldots\}$. Here, one is given a
metric space on $n$ points, and a parameter $k$, and the goal is now
to find a collection of $k$ subgraphs of type $X$ to cover all points
so that the maximum length of any of these subgraphs is smallest.  The
problem is APX-hard in the case of trees \cite{XW10} and
constant-factor approximation algorithms are known
\cite{AHL06,EG+04,KS14}. Xu, Xu and Li~\citeyear{XXL10} study the min-max
{\em path} cover problem and obtain constant-factor approximation
algorithms for several variants, also including delivery times.  

Another notable variant is the preemptive multi-vehicle dial-a-ride
problem, where $n$ items have to be transported by a fixed number of
vehicles, which are located at given depots.  Item $i\in\{1,\dots,n\}$
has to be picked up at  $s_i$ and delivered to $t_i$. Items may be passed
from one vehicle to another on their journey. G{\o}rtz, Nagarajan, and
Ravi \citeyear{GNR15} present an $\mathcal{O}(\log^3 n)$-approximation
algorithm for minimizing the makespan under capacity constraints, and
an $\mathcal{O}(\log t)$-approximation algorithm without vehicle
capacities, where $t$ is the number of distinct depots. 

%
%


\section{Deciding Feasibility}
\label{sec:feasibility}

\subsection{Notation}

Let us call an arborescence \emph{proper} if its root is $r$ and has out-degree 1, 
all elements of $P$ are vertices with out-degree 0 or 1,
and all other vertices have out-degree 2. 
We may restrict ourselves to schedules with proper arborescences because
if the root has out-degree 2 we can introduce an extra vertex at the same location without changing delay or cost. 

For a proper arborescence $(W,A)$ and $x\in W$ we use the following notation:
 $(W_{x*},A_{x*})$ is the maximal subarborescence of $(W,A)$ rooted at $x$.
 For $y\in W_{x*}$ we denote by $(W_{xy},A_{xy})$ the path from $x$ to $y$ in $(W,A)$.
 We have
 $W=W_0\cup W_1\cup W_2$, where $W_i$ contains the vertices of out-degree $i$ ($i=0,1,2$).
The elements of $W_0$ are called \textbf{leaves}, and the elements of $W_2$ are the \textbf{bifurcation nodes}. 
Note that $|W_0|=|W_2|+1$.

With this notation, we extend the definition of delay of an item in a schedule $(W,A,\mu)$ to any vertex $y \in W$:
$$\delay_{(W,A,\mu)}(r,y) \ := \ c(A_{ry},\mu) \, + \, \delta |P\cap W_{ry}| \, +  \!\sum_{w\in W_{ry}\cap W_2} \min_{(w,x)\in\delta^+(w)}  |P\cap W_{x*}|.$$
The first term is the delay of traversing edges, 
the second term is the time for delivering items on the $r$-$y$-path,
and the third term is the time to hand over a subset of items to a subtour.
Now, the delay of a schedule $(W,A,\mu)$ can be written as
$$\delay(W,A,\mu) \ := \ \max_{p\in P} \delay_{(W,A,\mu)}(r,p).$$
We denote by $n:=|P|$ the number of items.

\subsection{Caterpillar structure}



If we disregard cost, we can afford a separate vehicle for every item, going straight from the root to the item's destination.
This certainly minimizes the first two components (travel time and delivery time) of the delay to each item.
Then the only question is how the tree structure should look like, because it will determine the handover delays.
It turns out that there is always a fastest solution with a caterpillar structure. 
This allows us to determine efficiently whether a feasible solution exists.  

We first introduce the leafication step that takes a vertex with fanout one and branches it off
as a single leaf.
\begin{definition}[Leafication]
  Consider a  schedule $(W,A,\mu)$. Let $y \in W_1\setminus\{r\}$, and let $(x,y), (y,z)\in A$ be the two arcs incident to $y$ in $(W,A)$.
  The \textbf{leafication} of $y$ is the new schedule
  $(W', A', \mu')$ with $W' = W\dot{\cup}\{y'\}$, where $y'$ is a new vertex,
  $$A' \ = \ A \setminus \{(x,y), (y,z)\} \cup \left\{(x,y'), (y',y), (y',z)\right\},$$
  $\mu'(u) = \mu(u)$ for all $u \in W$, and $\mu'(y') = \mu(y)$. 
\end{definition}

\begin{figure}
\begin{center}
\includegraphics[width=0.35\columnwidth]{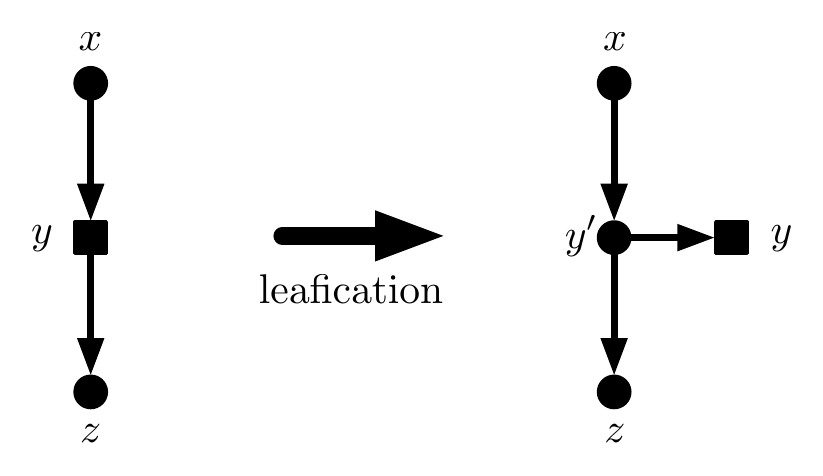}
\end{center}
\caption{\label{fig:leafication} Leafication of $y$.}
\end{figure}

See Figure \ref{fig:leafication}.
The leafication does not increase the delay of the schedule:

\begin{lemma}
  If 
  $(W',A',\mu')$ is obtained from $(W,A,\mu)$ through a leafication,
  then $$\delay(W',A',\mu') \ \le \ \delay(W,A,\mu).$$
\end{lemma}

\prove
  Let $y$ be the leafication vertex and $(x,y), (y,z) \in A$ its incident arcs.
  First note that the leafication changes delays only in $W_{y\star} =
  W'_{z\star} \cup \{y\}$. Furthermore,  $c(A_{rp},\mu) = c(A'_{rp},\mu')$ for all $p \in P$.
  Now for $y$ and any $w \in W_{z\star}\cap P$ we  have 
\begin{equation*}
\begin{array}{rcl}
\delay_{(W',A',\mu')}(r,y) &\le  &\delay_{(W',A',\mu')}(r,w) \\
                 & =   & \delay_{(W,A,\mu)}(r,w) - \delta + 1 \\
                 & \le & \delay_{(W,A,\mu)}(r,w) \\
                 & \le &\displaystyle  \max_{p\in P}\delay_{(W,A,\mu)}(r,p),
\end{array}
\end{equation*}
where the first inequality follows from the fact that $w$ collects at least the handover delay
as $y$ and at least its own delivery time.
The equality follows from replacing the delivery time by the handover time for $y$ on the path to $z$.
The second last inequality follows from $\delta\ge 1$. 
We conclude $\delay(W',A',\mu') \le \delay(W,A,\mu)$.
\endproof

By leafication we can get rid of out-degree 1 vertices (except for the root). In order to obtain the caterpillar structure we need to move all out-degree 2 vertices
onto a single ``heavy'' path:

\begin{definition}[Heavy path]
Given a proper arborescence $(W,A)$, a vertex $x\in W$ is called \textbf{heavy} if $x=r$ or 
$|W_{x*}\cap P| \ge |W_{y*}\cap P|$ for every child $y$ of the  predecessor of $x$. 
A \textbf{heavy path} in $(W,A)$ is maximal set of heavy vertices that induce a path.
\end{definition}

Note that any heavy path begins at the root and ends at a leaf.
We will move all bifurcation nodes onto a heavy path by the following operation.


\begin{definition}[Flip]
  Consider a  schedule $(W,A,\mu)$, and $H$ a heavy path in $(W,A)$. 
  Let $w \in W_2\cap H$, and let $h$ be the child of $w$ that belongs to $H$.
  Suppose that the other child $x$ of $w$ is a bifurcation node with $\mu(h)=\mu(x)$.
  Let $y_h$ and $y_l$ be the two children of $x$, where $y_h$ is heavy.
  The \textbf{flip} at $x$ is the new schedule $(W, A', \mu)$ with 
  $$A'\ := \ A \setminus \{(w,h),(x,y_l)\}\cup\{(w,y_l),(x,h)\}.$$
\end{definition}

\begin{wrapfigure}{R}{0.45\columnwidth}
\begin{center}
\includegraphics[width=0.45\columnwidth]{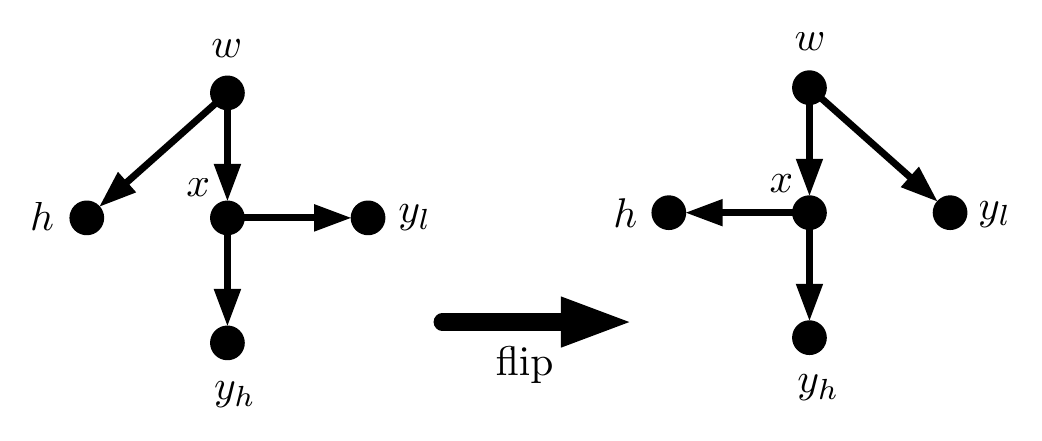}
\end{center}
\caption{\label{fig:flip-operation}Flip operation.}
\end{wrapfigure}

See Figure \ref{fig:flip-operation}.
Note that $H\cup\{x\}$ is a heavy path in $(W,A')$.
We now show that a flip does not make a schedule slower.

\begin{lemma}
  If $(W,A',\mu)$ is obtained from $(W,A,\mu)$ through a flip,
  then $$\delay(W',A',\mu') \ \le \ \delay(W,A,\mu).$$
\end{lemma}

\prove
First note that $c(A'_{rp},\mu) \le c(A_{rp},\mu')$ for all $p \in P$ due to $\mu(h)=\mu(x)$ and the triangle inequality. 
We show that the total handover delay to any item does not increase.
By construction the delays of items outside $W_{x*}$ coincide for $(W,A,\mu)$ and  $(W,A',\mu)$.
The handover delays of the flipped schedule are reduced by  $|W_{y_l*}\cap P|$ for all items in $W_{y_h*}$
and by  $|W_{x*}\cap P|$ for all items in $W_{y_l*}$.
\endproof

In graph theory, caterpillars are trees for which deleting all leaves results in a path. We use the term for arborescences in a slightly different way:

\begin{definition}[Caterpillar]
A proper arborescence $(W,A)$ is a \textbf{caterpillar} if $W_1=\{r\}$ and the subgraph induced by $\{r\}\cup W_2$ is a path.
For $P=\{p_1,\ldots,p_n\}$ we denote by $C(p_n,\ldots,p_1)$ the caterpillar in which the $r$-$p_i$-path in $(W,A)$ has $\min\{n,n+2-i\}$ edges for all $i=1,\ldots,n$.
\end{definition}

See Figure \ref{fig:reversecaterpillar}.
Now we can show the main result of this section.

\begin{theorem}
\label{thm:structure_of_fastest_solutions}
There exists a schedule $(W,A,\mu)$ with minimum delay
such that $(W,A)$ is a caterpillar and $\mu(w)=\mu(r)$ for all $w\in W_2$.
\end{theorem}


\prove
If $n = 1$, the statement is clearly true, so let $n> 1$. 
Let $(A^{0}, W^{0}, \mu^{0})$ be a schedule with minimum delay.
Recall that $W^0_0 \cup W^0_1 = \{r\} \cup P$. 

\caseclaim{Step 1} 
By iteratively applying a leafication step to all $y \in W^0_1\setminus\{r\}$, we can transform it into a schedule
$(A^{1}, W^{1}, \mu^{1})$ with $W^{1}_1=\{r\}$ that still has minimum delay.
Then all items are delivered at leaves, i.e.,  $P = W^1_0$.

\caseclaim{Step 2}  
We transform  $(A^{1}, W^{1}, \mu^{1})$ into $(A^{2}, W^{2}, \mu^{2})$, by setting $A^2=A^1, W^2=W^1$
and $\mu^2(y) = \mu^1(r) $ for all $y \in W^2_2$.
As $(A^{1}, W^{1}) = (A^{2}, W^{2})$, only the delays for traversing arcs change in $(A^{2}, W^{2}, \mu^{2})$.
But as $c(A^2_{rp},\mu^2) = c(r,p)$ for all $p\in P$, this part of the delay is minimum
possible and so $(A^{2}, W^{2}, \mu^{2})$ still has minimum delay.

\caseclaim{Step 3} 
Finally, we use the flip operation to obtain the caterpillar structure. Let $H$ be a heavy path.
As long as there is a bifurcation node outside $H$, there is such a vertex $x$ such that its predecessor $w$ belongs to $H$.
Then applying the flip operation at $x$ increases the cardinality of $H$ by one, so after finitely many steps, the
heavy path contains all bifurcation nodes.
\endproof

\subsection{Consequences}

\begin{wrapfigure}{R}{0.4\columnwidth}
\begin{center}
\includegraphics[width=0.4\columnwidth]{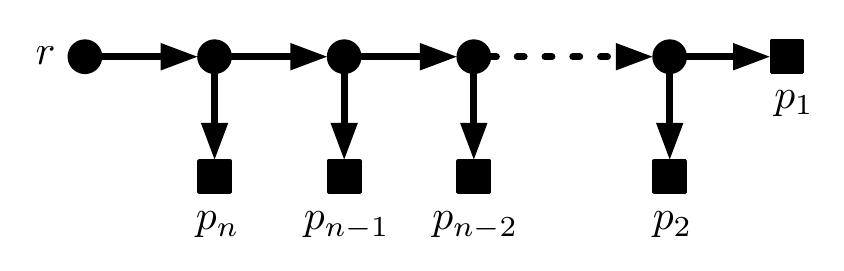}
\end{center}
\caption{\label{fig:reversecaterpillar}Caterpillar $C(p_n,\ldots,p_1)$.} 
\end{wrapfigure}

\begin{corollary}
\label{boundsonDelta}
For any feasible instance we have
\begin{enumerate}[(a)]
\item $\Delta\ge \min \bigl\{c(r,q) + \delta + \min\{|Q|,n-1\} : q\in Q \bigr\}$ for every nonempty subset $Q\subseteq P$;
\item  $\Delta\ge n$;
\item $\Delta\ge \max\{c(r,p):p\in P\}$.
\end{enumerate}
\end{corollary}

\prove
For (a), let $Q\subseteq P$, and consider a feasible caterpillar (which exists by Theorem \ref{thm:structure_of_fastest_solutions}).
At least one of the $|Q|$ items, say $q$, will have at least $\min\{|Q|,n-1\}$ bifurcation nodes on its path.
The path to $q$ has length $c(r,q)$ and it also pays $\delta$ for delivering $q$.

(b) follows from (a) by setting $Q=P$ and using $\delta\ge 1$.
(c) is obvious.
 \endproof

Theorem~\ref{thm:structure_of_fastest_solutions} allows us to determine efficiently
whether a feasible  schedule exists.

\begin{corollary}\label{cor:fastest_solutions}
  \label{cor:fastest_schedule}
  We can find a schedule meeting the deadline or decide that none exists 
  in time $\mathcal{O}(n\log n + \theta n)$, 
  where $\theta$ is the time to evaluate distances in $(M,c)$.
\end{corollary}

\prove
If $n=1$, verifying the feasibility of the deadline is easy, so let $n> 1$.
By Theorem~\ref{thm:structure_of_fastest_solutions} there is a  fastest solution $(A,W,\mu)$
such that $(A,W)$ is a caterpillar.
This caterpillar is unique up to the order in which the items are attached as leaves. Furthermore,  
the distribution of total handover delays for the leaves is predetermined  
and each item suffers a single delivery delay of $\delta$ for its own delivery.
So the leaves have delivery plus handoff delay
$1+\delta,2+\delta,\ldots,n-2+\delta,n-1+\delta,n-1+\delta$.

Thus it suffices to find an assignment of the items to the leaves
that meets the deadline.  The best we can do is to iteratively assign
an item with the maximum distance from $r$ to the closest available leaf
in the arborescence.  To this end we sort the items by their
distance from $r$.  Let $p_1,p_2,\dots,p_{n} $ be the ordering of
the items by non-decreasing distance from $r$, i.e. $c(r,p_i) \le c(r,p_{i+1})$ for all $i \in \{1,\dots,n-1\}$.

We assign the items $p_1, p_2, \dots, p_{n}$ one by one to a not yet occupied leaf
vertex that has a maximum number of
arcs on its path from $r$ (cf.\ Figure \ref{fig:reversecaterpillar}).
The deadline $\Delta$ can be met if and only if the generated schedule meets it.

The running time is dominated by sorting the items, which can be done in $\mathcal{O}(n\log n)$ time,
and by computing all root to item distances, which takes $\theta\cdot n$ time.
\endproof

The schedule from Theorem~\ref{thm:structure_of_fastest_solutions} is fast but also 
expensive. It has the maximum possible setup cost of $\sigma\cdot n$ and also high travel costs, as each item is transported individually
from the root location to its destination.

\section{Algorithm}
\label{sec:algorithm}

%

Our algorithm first groups the items by splitting a short tour into paths similar to \cite{AHL06}.
The items in each group will have similar distance from $r$. 
Therefore, rearranging the fastest solution with the caterpillar structure (Theorem \ref{thm:structure_of_fastest_solutions})
so that the items in each group are consecutive
does not make the schedule much slower.
Next, we design a two-level caterpillar, where the items in each group are served by a caterpillar and the groups are
served by a top-level caterpillar. 
In order to avoid that all subtours begin at the position of the root, we make the main tour of each sub-caterpillar
drive to all locations of items in that group.
Finally, we avoid too many subtours by merging tours in each subcaterpillar.

\subsection{Grouping items}
\label{sec:nearearly}

\begin{lemma}
\label{nearearlyseparatepaths}
Let $\mst$ denote the length of a minimum cost tree with vertex set $\{r\}\cup P$,
and let $s\in P$ be an item at maximum distance from $r$. 
Let $\epsilon>0$ and $0<\Delta\le\mst$.
Then there is a forest $(P,F)$ whose components are vertex-disjoint paths 
 such that 
\begin{enumerate}[(a)]
\item the number of paths is at most $1+\frac{n+2\,\mst-c(r,s)}{\epsilon\Delta}$;
\item no path is longer than $\epsilon\Delta$;
\item no path contains more than $1+\epsilon\Delta$ items;
\item $c(F)\le 2\, \mst - c(r,s)$.
\end{enumerate}
Such a forest can be found in polynomial time.
\end{lemma}

\prove
Take any approximately cheapest path with vertex set $\{r\}\cup P$ from $r$ to $s$.
We can find it by taking a minimum cost spanning tree for $\{r\}\cup P$ in $(M,c)$,
doubling all edges except those of the $r$-$s$-path, finding an Eulerian $r$-$s$-walk, and shortcutting.
The resulting $r$-$s$-path $(\{r\}\cup P,F_0)$ has total cost at most $2 \,\mst-c(r,s)$.
From now on, we will only delete edges, yielding (d).

To satisfy (b) and (c), we start with $F=\emptyset$ and traverse the path $r$-$s$-path, ignoring $r$ and the first edge. 
In each step, we add the next edge to $F$ unless this would violate (b) or (c). 

The conditions (b), (c), and (d) are then satisfied by construction.
Whenever we drop an edge $e$, one of the conditions (b) and (c) would be violated for $P\cup\{e\}$, 
where $P$ is a connected component of $F$. 
If (b), the length of $P\cup\{e\}$ exceeds $\epsilon\Delta$, so this can happen at most $\frac{c(F_0)}{\epsilon\Delta}$ times.
If (c), the number of items in $P$ exceeds $\epsilon\Delta$, so this can happen at most $\frac{n}{\epsilon\Delta}$ times. 
So we drop at most  $\frac{2\,\mst-c(r,s)}{\epsilon\Delta} + \frac{n}{\epsilon\Delta}$ edges (in addition to the initial one). This yields (a).
 \endproof

Note that (b) immediately implies that items in the same path have similar distances from $r$:
if $p$ and $p'$ are in the same path, the triangle inequality yields $c(r,p) \le c(r,p')+\epsilon\Delta$.

\subsection{Towards a cheaper schedule}

Let us call the vertex sets of the connected components of $(P,F)$ from Lemma~\ref{nearearlyseparatepaths} {\em groups}; they form a partition of $P$.
For $p\in P$, let 
$$d(p):=\max\{c(r,p'): \text{$p$ and $p'$ are in the same group}\}$$
be the distance from $r$ to the most remote item in the same group as $p$ (note that $p'=p$ is not excluded). 
Order the items $P=\{p_1,\ldots,p_n\}$ such that 
$$d(p_1)\le \cdots \le d(p_n)$$
and 
$$F\subseteq \{\{p_{i},p_{i+1}\}:i=1,\ldots,n-1\},$$
so items of the same group are consecutive, every edge in $F$ connects two consecutive items, and groups containing more remote items come later.

\begin{figure}
\begin{center}
\includegraphics[width=0.95\columnwidth]{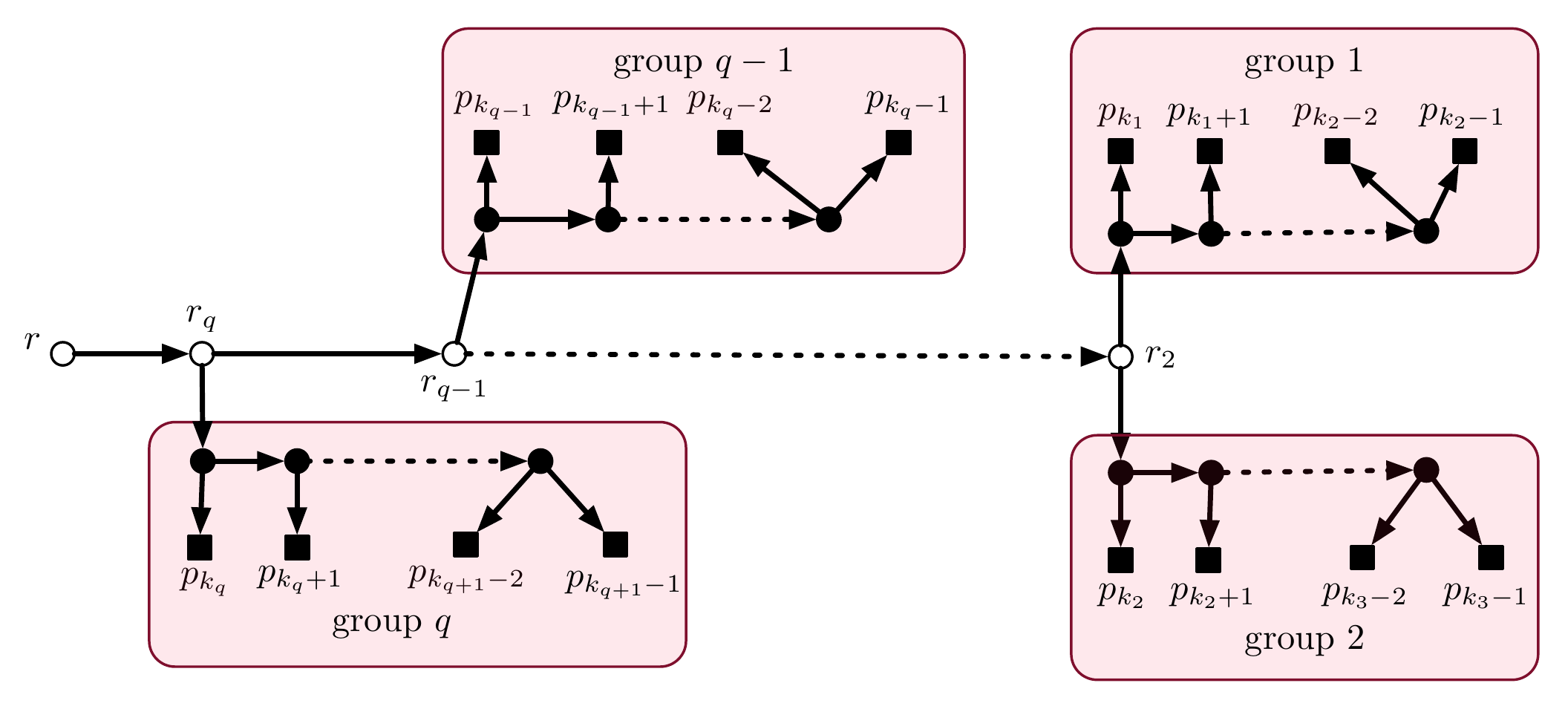}
\end{center}

\caption{\label{fig:cat2} Schedule $S_1$ with a sub-caterpillar for each group.}
\end{figure}

Let $P=\{p_1,\ldots,p_n\}$ be the items ordered as above, and let $1=k_1<k_2<\cdots<k_{q+1}=n+1$ such that
$\{p_{k_i},\ldots,p_{k_{i+1}-1}\}$ ($i=1,\ldots,q$) are the groups.
Let $S_1$ be the schedule resulting from a path with vertices $r,r_q,r_{q-1},\ldots,r_2$ in this order
by identifying the root of the caterpillar $C(p_{k_i},\ldots,p_{k_{i+1}-1})$ with $r_{\max\{2,i\}}$ ($i=1,\ldots,q$; see Figure \ref{fig:cat2}).
The bifurcation nodes $r_q,\ldots,r_2$ are placed at $\mu(r)$, but the bifurcation nodes 
of the subcaterpillars are placed at the position of the item that splits off there:
the $j$-th bifurcation node of the subcaterpillar for group $i$ is placed at $\mu(p_{k_i+j-1})$.

\begin{lemma}
\label{subcaterpillars2} 
If the instance is feasible, then $\delay(S_1)\le (1+3\epsilon)\Delta$. 
\end{lemma} 

\prove
Consider group $i$ with items $p_{k_i},\ldots,p_{k_{i+1}-1}$.
The maximum delay of an item in this group is the one to $p_{k_{i+1}-1}$, which is at most
$$c(r,p_{k_i})+\sum_{l=k_i}^{k_{i+1}-2} c(p_l,p_{l+1}) + n - k_{\max\{2,i\}} + 1 + k_{i+1} - k_i - 1 + \delta,$$
which by Lemma \ref{nearearlyseparatepaths}(b) and (c) is at most
\begin{equation}
\label{boundongroupdelay}
d(p_{k_i})+\epsilon\Delta + n - k_{\max\{2,i\}} + 1 + \epsilon\Delta + \delta.
\end{equation}
By Corollary~\ref{boundsonDelta}~(a), there exists a $j\in\{k_i,\ldots,n\}$ with 
$\Delta\ge c(r,p_j) + n - \max\{2,k_i\} + 1 + \delta$.

Since $d(p_{k_i}) \le d(p_j) \le c(r,p_j) + \epsilon\Delta$,
we have 
\begin{eqnarray*}
\Delta &\ge& c(r,p_j) + n - \max\{2,k_i\} + 1 + \delta \\
&\ge& c(r,p_j) + n - k_{\max\{2,i\}} + 1 + \delta \\
&\ge& d(p_{k_i})-\epsilon\Delta + n - k_{\max\{2,i\}} + 1 + \delta.
\end{eqnarray*}
Hence our bound \eqref{boundongroupdelay} on the maximum delay of an item in group $i$
yields that this delay is at most $(1+3\epsilon)\Delta$.
%
%
\endproof

We can bound the length of $S_1$ as follows:

\begin{lemma}\label{lem:len}
$S_1$ has length  at most $(2 + \frac{2}{\epsilon})\mst$.
\end{lemma}

\prove
The length of the schedule is $\sum_{i=1}^q c(r,p_{k_i})$ (which is the initial edges of the sub-caterpillars) plus $c(F)$
(remaining length of the subcaterpillars). 
By Lemma \ref{nearearlyseparatepaths}, $c(F)\le 2\,\mst - c(r,s)$
and the number $q$ of groups is at most $1+ \frac{n+2\,\mst - c(r,s)}{\epsilon\Delta}$.
Moreover $c(r,p_{k_i})\le c(r,s)$ for all $i$ by the choice of $s$.
Hence the length of the schedule is at most
$$    q \, c(r,s) + 2\,\mst - c(r,s) \le 
    \frac{n+2\,\mst - c(r,s)}{\epsilon\Delta} \, c(r,s) + 2\,\mst.
$$
Since $n\le\Delta$ 
and
$c(r,s)\le \Delta$ by Corollary~\ref{boundsonDelta}~(b) and (c), this is at most
$$
\frac{\Delta}{\epsilon\Delta} c(r,s) +
  \frac{2\,\mst - c(r,s)}{\epsilon\Delta} \Delta + 2\,\mst 
=  \left(2 + \frac{2}{\epsilon}\right) \,\mst.
$$
\endproof

\subsection{Saving vehicles}

The schedule is already short, but it still contains a separate vehicle for each item.
However, we can deliver up to $m:=1+\lfloor\frac{\epsilon\Delta}{\delta}\rfloor$ items by the same vehicle
by replacing the edge entering an item $p_j$ in group $i$ by the edge $(p_{j-1},p_j)$ unless $j-k_i$ is a multiple of $m$.
The resulting schedule $S_2$ is the output of our algorithm, except that we remove the non-item nodes that now have out-degree 1, by shortcutting.

\begin{lemma}
\label{finalbounds}
$S_2$ has delay at most $(1+4\epsilon)\Delta$ and
length at most $(4 + \frac{2}{\epsilon})\mst$. It has at most $1+ \frac{2}{\epsilon} \left(\frac{\mst + n\delta}{\Delta} \right)$ vehicles.
\end{lemma}

\prove
Going from $S_1$ to $S_2$ increases the maximum delay of an item in a group by at most $(m-1)\delta \le \epsilon\Delta$,
because the maximum travel time and the maximum handover delay in each group cannot increase.

The length of $S_2$ is at most $c(F)$ longer than $S_1$. Since $c(F)\le 2\mst$, the length bound follows.

The total number of vehicles is at most $$q+ \frac{n}{m} \le q+\frac{\delta n}{\epsilon\Delta}\le 
1+ \frac{n + 2\,\mst - c(r,s)}{\epsilon\Delta} + \frac{\delta n}{\epsilon\Delta} 
\le  1+ \frac{2}{\epsilon} \left(\frac{\mst + n\delta}{\Delta} \right),$$
where we used $\delta\ge 1$ in the last inequality.
\endproof

%
%

We compare this to the following lower bound. 
\begin{lemma}
\label{lem:lowerboundonnumberoftours}
Every feasible schedule has length at least $\frac{1}{2}\mst$ and uses at least $\frac{\frac{1}{2}\text{\scriptsize \mst}+n\delta}{\Delta}$ vehicles,
where $\mst$ is the length of a minimum spanning tree for $\mu(\{r\}\cup P)$.
\end{lemma}

\prove
Any schedule connects $\{r\}\cup P$, so the lower bound $\frac{1}{2}\mst$ for the length follows from the Steiner ratio.
For the bound on the number of vehicles, fix any feasible schedule $(W^*,A^*,\mu^*)$, say with $l^*$ vehicles numbered $1,\ldots,l^*$. 
Let $D_i$ be the delay of the last item that vehicle $i$ delivers; this is at most $\Delta$. 
As each edge needs to be traversed and each item delivered by one of the vehicles,
\begin{equation*}
\frac{1}{2} \mst+n\delta \ \le \ c(A^*,\mu^*)+n\delta \ \le \ \sum_{i=1}^{l^*} D_i \ \le \ l^*\Delta.
\end{equation*}
\mathendproof

Lemma \ref{finalbounds} and \ref{lem:lowerboundonnumberoftours} imply that
the cost of our schedule is at most $8+\frac{4}{\epsilon}$ times the cost of an optimum schedule.
This proves Theorem \ref{main}.

Our algorithm is very fast -- the running time is dominated by computing a minimum spanning tree and sorting the groups.

We remark that the constants can be improved, for example by beginning with a Steiner tree that is a better approximation than MST.
However, any improvement by more than a constant factor would imply an improvement over the 25-year old bicriteria algorithm 
for shallow-light trees by Cong et al. \citeyear{CK+92}. We will demonstrate this in the following.

\section{An almost tight example}
\label{sec:example}

We now  show that our bicriteria result
is the best we can hope for up to constant factors.
To this end, we modify an example of Khuller, Raghavachari, and Young~\citeyear{KRY95} to make it work for a uniform deadline.

\begin{theorem}
Let $\epsilon >0$ and  $1 \le  \alpha < 1+\frac{1}{\epsilon}$.
There is an undirected graph $G=(V,E)$ with weights $c:E\to\mathbb{R}_{>0}$ and $r\in V$
such that $\mathrm{dist}_{(G,c)}(r,v) \le 1$ for all $v \in V$, 
and for each spanning tree $F$ in which every path from $r$ has length at most $1+\epsilon$, the total length of $F$ is more than
$\alpha \cdot \mst$, where $\mst$ is the length of a minimum spanning tree.
\end{theorem}

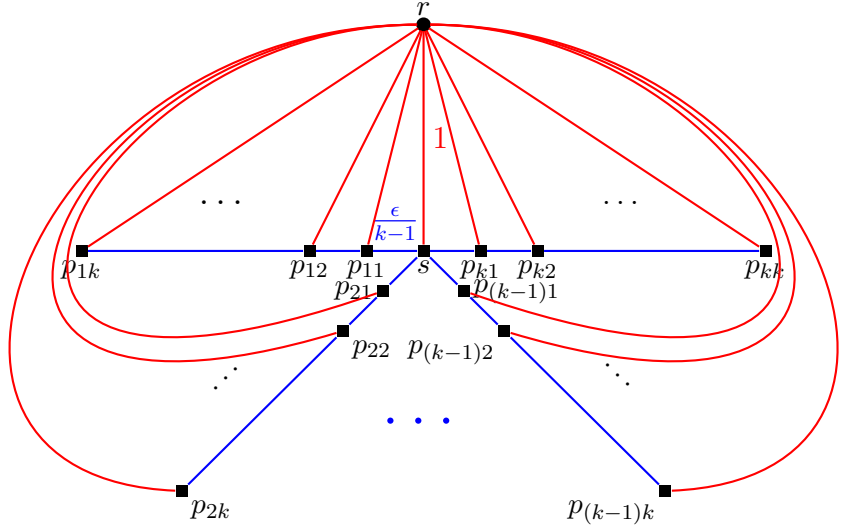
\begin{figure}[!t]

\begin{tikzpicture}[scale=1.5, very thick, minimum size = 5, inner sep = 3,square/.style={regular polygon,regular polygon sides=4}]
\pgfmathsetmacro {\xa }{ cos (45)}
\pgfmathsetmacro {\ya }{ sin (45)}
\node[draw,circle,fill, scale=0.5] (r) at (0,2) {};       \node[above]  at (r)  {\small $r$};
\node[fill,square, scale =0.5] (v) at (0,0) {};\node[below]  at (v)  {\small $s$};

\node[fill,square, scale =0.5] (p11) at (-0.5,0) {};\node[below]  at (p11)  {\small $p_{11}$};
\node[fill,square, scale =0.5] (p12) at (-1,0) {};\node[below]  at (p12)  {\small $p_{12}$};
\node[] (p13) at (-1.75,0) {};\node[above=0.5cm]  at (p13)  {\large $\dots$};
\node[fill,square, scale =0.5] (p1k) at (-3,0) {};\node[below]  at (p1k)  {\small $p_{1k}$};

\node[fill,square, scale =0.5] (pk1) at (0.5,0) {};\node[below]  at (pk1)  {\small $p_{k1}$};
\node[fill,square, scale =0.5] (pk2) at (1,0) {};\node[below]  at (pk2)  {\small $p_{k2}$};
\node[] (pk3) at (1.75,0) {};\node[above=0.5cm]  at (pk3)  { $\dots$};
\node[fill,square, scale =0.5] (pkk) at (3,0) {};\node[below]  at (pkk)  {\small $p_{kk}$};

\node[fill,square, scale =0.5] (p21) at (-0.5*\xa, -0.5*\ya) {};\node[left]  at (p21)  {\small $p_{21}$};
\node[fill,square, scale =0.5] (p22) at (-1*\xa,-1*\ya) {};\node[below right]  at (p22)  {\small $p_{22}$};
\node[] (p23) at (-2*\xa,-2*\ya) {};\node[rotate = 45, above=0.5cm]  at (p23)  { $\dots$};
\node[fill,square, scale =0.5] (p2k) at (-3*\xa,-3*\ya) {};\node[below right]  at (p2k)  {\small $p_{2k}$};

\node[fill,square, scale =0.5] (pk-1) at (0.5*\xa, -0.5*\ya) {};\node[right]  at (pk-1)  {\small $p_{(k-1)1}$};
\node[fill,square, scale =0.5] (pk-2) at (1*\xa,-1*\ya) {};\node[below left]  at (pk-2)  {\small $p_{(k-1)2}$};
\node[] (pk-3) at (2*\xa,-2*\ya) {};\node[rotate = -45, above=0.5cm]  at (pk-3)  { $\dots$};
\node[fill,square, scale =0.5] (pk-k) at (3*\xa,-3*\ya) {};\node[below left]  at (pk-k)  {\small $p_{(k-1)k}$};

\node[blue,scale=2] (bluedots) at (0,-1.5) {\dots};

\draw[red,thick] (r)-- node[right] {$1$} (v);

\draw[blue,thick] (v)--node[above]{$\frac{\epsilon}{k-1}$}(p11);
\draw[blue,thick] (p11)--(p12);
\draw[blue,thick] (p12)--(p1k);

\draw[blue,thick] (v)--(p21);
\draw[blue,thick] (p21)--(p22);
\draw[blue,thick] (p22)--(p2k);

\draw[blue,thick] (v)--(pk-1);
\draw[blue,thick] (pk-1)--(pk-2);
\draw[blue,thick] (pk-2)--(pk-k);

\draw[blue,thick] (v)--(pk1);
\draw[blue,thick] (pk1)--(pk2);
\draw[blue,thick] (pk2)--(pkk);

\draw[red,thick] (r) -- (p11);
\draw[red,thick] (r) -- (p12);
\draw[red,thick] (r) -- (p1k);

\draw[red,thick] (r) -- (pk1);
\draw[red,thick] (r) -- (pk2);
\draw[red,thick] (r) -- (pkk);

\draw[red,thick] (r) ..controls (-3,2) and  (-5,-2).. (p21);
\draw[red,thick] (r) ..controls (-3.2,2) and  (-5,-2).. (p22);
\draw[red,thick] (r) ..controls (-3.4,2) and  (-5,-2)..(p2k);

\draw[red,thick] (r) ..controls (3,2) and (5,-2)..(pk-1);
\draw[red,thick] (r) ..controls (3.2,2) and (5,-2)..(pk-2);
\draw[red,thick] (r) ..controls (3.4,2) and (5,-2)..(pk-k);
\end{tikzpicture}

\caption{An almost tight example}
\label{fig:almost-tight-example}
\end{figure}

\prove
For sufficiently large  $k\in \Nset$ (in particular $k>1+\epsilon$), consider the 
graph $G=(V,E)$ with vertex set $V=\{r,s,p_{11},\dots,p_{1k},p_{21},\dots, p_{2k},\dots, p_{k1},\dots, p_{kk}\}$  
shown in Figure~\ref{fig:almost-tight-example}.  
The edge set $E$ contains a red edge from $r$ to every vertex other
than $r$, and 
blue edges $\{s,p_{i1}\}$ and $\{p_{ij},p_{i(j+1)}\}$ for all $i=1,\dots,k$ and  $j=1,\dots, k-1$.
Red edges have weight 1, and blue edges have weight $\frac{\epsilon}{k-1}$.
Thus, $\mathrm{dist}_{(G,c)}(r,v) =1$ for all $v \in V\setminus\{r\}$.

For each $1\le i\le k$, unless the tree uses one of the red edges $\{r,p_{ij}\}$ ($j=1,\dots k$),
the tree distance from $r$ to $p_{ik}$ is at least $1+\frac{\epsilon}{k-1}k > 1+\epsilon$.
Therefore, for any tree $F$ in which every path from $r$ has length at most $1+\epsilon$, $F$ has
total length at least 
$$k + k(k-1)\frac{\epsilon}{k-1} + \frac{\epsilon}{k-1} > k (1+\epsilon).$$

On the other hand, a minimum spanning tree consists of one red and all blue edges; it has length $\mst = 1+k^2\frac{\epsilon}{k-1}$.
Thus, the ratio between the total length of a tree whose paths from $r$ have length at most  $1+\epsilon$ and a minimum spanning tree is at least
$$
\frac{k(1+\epsilon)}{1+k^2\frac{\epsilon}{k-1}}   \xrightarrow{k\to\infty} \frac{1+\epsilon}{\epsilon} = 1+\frac{1}{\epsilon} > \alpha.
$$
So for sufficiently large $k$, the ratio is greater than $\alpha$.
\endproof

This shows that the result of Cong et al.~\citeyear{CK+92} mentioned in Section \ref{sec:relatedwork} is best possible
up to a constant factor.

This example applies not only to shallow-light trees, but also to a special case of our problem, namely when
$M=\mu(\{r\}\cup P)$, $\sigma=0$, and $c$ is large compared to $\delta$ so that delivery times and handover delays can be neglected.
We see that unless we use a much stronger lower bound than $\mst$ on the length of a feasible schedule, the tradeoff in Theorem \ref{main} is unavoidable.

\bibliographystyle{plainnat}
\bibliography{parceldelivery.bib}

\begin{appendices}

\section{APX-hardness of TSP with one fixed endpoint}
\label{sec:tsppathapxhard}

It is well-known that the metric TSP is APX-hard \cite{PY93}. More precisely, it is NP-hard to approximate it with any ratio better than $1+\epsilon$, where $\epsilon=\frac{1}{122}$ \cite{KLS15}.
We deduce from this inapproximability thresholds for the variants studied by Hoogeveen \citeyear{HO91}, where we look for a path with 0, 1, or 2 endpoints fixed.

First, the same threshold holds for the path TSP if both endpoints are fixed. Given a TSP instance, just guess an edge $\{r,s\}$ of an optimal tour and approximate the $r$-$s$-path TSP instance.

Second, we get the threshold $1+\frac{\epsilon}{3}$ if only one endpoint $r$ is fixed. This works as follows.
Given an instance of the path TSP with both endpoints $r$ and $s$ fixed, let $U$ be an upper bound on the optimum, say at most $\frac{5}{3}\opt$.
Add a vertex $t$ with distances $c(v,t):=c(v,s)+M$ for all cities $v$, where $M=\frac{3+\epsilon}{3-\epsilon} U$.
Then consider the path TSP with only one endpoint $r$ fixed.
The optimum has length at most $\opt+M$. 
In fact equal to this, because any tour not ending in $t$ will have length more than $2M =(1+\frac{\epsilon}{3})M+(1-\frac{\epsilon}{3})\frac{3+\epsilon}{3-\epsilon} U \ge (1+\frac{\epsilon}{3})(M+\opt)$.
Therefore, any algorithm with approximation ratio less than $1+\frac{\epsilon}{3}$ will find a tour ending in $t$ and have length less than
$(1+\frac{\epsilon}{3}) (\opt+M)$. Without loss of generality it visits $s$ just before $t$ (it is on the way anyway).
After deleting the edge $\{s,t\}$ we get a tour for the original $r$-$s$-path TSP instance, and it has length less than
$(1+\frac{\epsilon}{3})(\opt+M) -M = (1+\frac{\epsilon}{3})\opt +\frac{\epsilon}{3} \frac{3+\epsilon}{3-\epsilon} U <(1+\frac{\epsilon}{3})\opt + \frac{\epsilon}{3}\cdot\frac{6}{5} U \le (1+\epsilon)\opt$,
where we used $\epsilon <\frac{1}{11}$ in the strict inequality.
 
Third, if no endpoint is fixed, we can apply the same trick twice, appending $q$ at $r$ and (as before) $t$ at $s$, and forcing any cheap path to be a path from $q$ to $t$.

\section{Steiner nodes}
\label{steiner-appendix}

\begin{figure}[!t]
\begin{center}
\begin{tikzpicture}[scale=1, very thick, minimum size = 5, inner sep = 1]
\node[fill,circle] (r) at (0,0) {}; \node at (0,0.3) {$r$};
\node[fill,circle] (s) at (10,0) {}; \node at (10,0.3) {$s$};
\node[fill] (p1) at (11,1) {}; \node at (11.5,1) {$p_1$};
\node[fill] (p2) at (11,0.5) {}; \node at (11.5,0.5) {$p_2$};
\node[fill] (pn) at (11,-1) {}; \node at (11.5,-1) {$p_n$};
\draw (r)--(s);
\draw (s)--(p1);
\draw (s)--(p2);
\draw (s)--(pn);
\node at (5,-0.3) {\small $n^2$};
\node at (9.8,-0.7) {\small $\frac{\epsilon n^2+2n}{2-\epsilon}$};
\draw[dotted] (11,0) -- (11,-0.5);
\end{tikzpicture}
\end{center}
\caption{An example where using a Steiner node reduces the cost very much.}
\label{fig:steiner-example}
\end{figure}
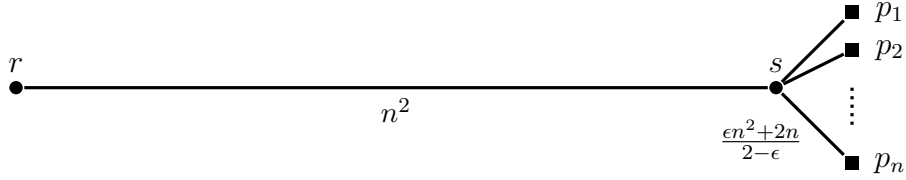

Our model allows to place bifurcation nodes at positions in $M\setminus(\mu(\{r\}\cup P)$, but our algorithm does not do this.
We remark that using such Steiner nodes would also be necessary to improve on Theorem \ref{main} by more than a constant factor.
Let $n\in\mathbb{N}$, $0<\epsilon<\frac{1}{2}$, and 
$M:=\{r,s,p_1,\ldots,p_n\}$ with $\mu(v)=v$ for $v\in\{r,p_1,\ldots,p_n\}$ and
$c(r,s)=n^2$ and $\frac{1}{2}c(p_i,p_j)=c(s,p_i)=\frac{\epsilon n^2 + 2n}{2-\epsilon} = c(r,p_i)-n^2$ for all $1\le i,j \le n$
(cf.\ Figure \ref{fig:steiner-example}).
Let $\sigma=0$, $\delta=1$ and $\Delta=n^2+n+\frac{\epsilon n^2 + 2n}{2-\epsilon} = \frac{2n^2 +4n -\epsilon n}{2-\epsilon}$.
Then traveling first to $p_i$ and then to $p_j$ ($j\not=i$) takes time
$\Delta - n + \frac{2(\epsilon n^2+2n)}{2-\epsilon}
= \Delta  + \frac{\epsilon(2n^2+n)+2n}{2-\epsilon}
> \Delta  + \frac{\epsilon(2n^2+4n -\epsilon n)}{2-\epsilon} = (1+\epsilon)\Delta$.
Unless we allow violating the deadline by more than a factor $1+\epsilon$, no tour can deliver more than one item, and no subtour can
start at any $p_i$. Hence either $n$ tours start at $r$, or subtours start at $s$, which decreases the cost by a factor
$$\frac{n(n^2+\frac{\epsilon n^2+2n}{2-\epsilon})}{n^2+n(\frac{\epsilon n^2+2n}{2-\epsilon})}
\ = \ \frac{2n^3+2n^2}{\epsilon n^3+(4-\epsilon)n^2}
\ \ \xrightarrow[n\to\infty] \ \ \frac{2}{\epsilon}.$$ 
 
\end{appendices}

\end{document}